\documentclass[superscriptaddress, prl, twocolumn]{revtex4-1}

\usepackage{times}
\usepackage{graphicx}
\usepackage{subfigure}
\usepackage{dcolumn}
\usepackage{bm}
\usepackage{multirow}
\usepackage{cancel}

\def\ket#1{\mathinner{|{#1}\rangle}}

\makeatletter
\def\Ddots{\mathinner{\mkern1mu\raise\p@
\vbox{\kern7\p@\hbox{.}}\mkern2mu
\raise4\p@\hbox{.}\mkern2mu\raise7\p@\hbox{.}\mkern1mu}}
\makeatother
\usepackage{amsmath}
\usepackage{amsthm}
\usepackage{amssymb}
\usepackage{rotating}
\usepackage{epstopdf}
\usepackage{bbold}
\newcommand{\sz}[0]{\ensuremath{\mathbf{\sigma}_z}}
\newcommand{\sx}[0]{\ensuremath{\mathbf{\sigma}_x}}
\newcommand{\sy}[0]{\ensuremath{\mathbf{\sigma}_y}}
\renewcommand{\sp}[0]{\ensuremath{\mathbf{\sigma}_{+}}}
\newcommand{\sm}[0]{\ensuremath{\mathbf{\sigma}_{-}}}
\usepackage{color}

\begin{document}
\author{Yuting Ping}
\email{yuting.ping@materials.ox.ac.uk}
\affiliation{Department of Materials, University of Oxford, Oxford OX1 3PH, United Kingdom}
\author{John H. Jefferson}
\affiliation{Department of Physics, Lancaster University, Lancaster LA1 4YB, United Kingdom}
\author{Brendon W. Lovett}
\email{b.lovett@hw.ac.uk}
\affiliation{School of Engineering and Physical Sciences, Heriot-Watt University, Edinburgh EH14 4AS, United Kingdom}
\affiliation{Department of Materials, University of Oxford, Oxford OX1 3PH, United Kingdom}

\title{A coherent and passive one dimensional quantum memory}

\begin{abstract}
We show that the state of a flying qubit may be transferred to a chain of identical, (near) ferromagnetically polarised, but non-interacting, static spin-$\frac{1}{2}$ particles in a {\it passive} way. During this process the flying qubit is coherently polarised, emerging in the direction of the majority static spins. We also show that this process is reversible for at least two flying qubits injected sequentially and thus has the potential to be exploited as a passive quantum memory to encode the flying qubits {\it without} the necessity of resetting between successive encoding operations. We show that the quantum information may be spread over many static spins in the memory chain, making the mechanism resistent to spin decoherence and other imperfections. Among some potential architectures, we discuss implementing the memory in a photonic waveguide embedded with quantum dots, which is resilient to various possible errors.
\end{abstract}

\maketitle
Robust quantum state transfer ({\it QST}) plays an important role in the field of quantum information processing ({\it QIP}), achieving quantum transmission of data through space or time~\cite{nielsen00}. Over the past decade numerous efforts have been made in this area, and many potentially feasible approaches have been suggested for both state transportation~\cite{yao11a} and storage~\cite{morton08, wesenberg09} in a variety of physical systems. However, in most instances, exercising the required level of active quantum control remains a challenging aspect of the current technology: errors are most likely introduced. On the other hand, it has recently been proposed that schemes utilising iterative applications of quantum maps can perform certain QIP tasks with reduced level of quantum control~\cite{ciccarello10, ping11}.

\begin{figure}[b]
\begin{center}
\includegraphics[width=.5\textwidth]{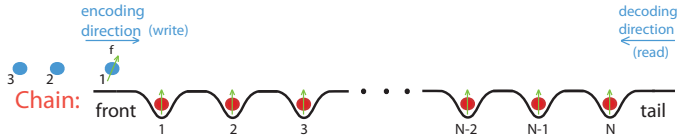}
\caption{Illustration of the passive memory system (not to scale) with the associated QST scheme. The coherent quantum memory consists of a sufficiently long ferromagnetic chain $\ket{F}$ of non-interacting, static spins (red). Each flying qubit (blue) enters the front end of the chain, interacts with the static spins sequentially and eventually emerges as polarised $\ket{\uparrow}_f$ at the tail end. The chain can then encode subsequent flying qubits sequentially in the same fashion. To read out from the memory, one simply injects a polarised qubit $\ket{\uparrow}_f$ back from the tail, and the state of the last encoded flying qubit is then recovered automatically out from the front. More states can be recovered sequentially in the same way by further injecting back polarised qubits $\ket{\uparrow}_f$ from the tail. There is no need to reset between successive encoding (decoding) rounds. Any multi-partite entanglement between injected qubits is also recovered in the read operations.}
\label{fig:device}
\end{center}
\end{figure}

In this Letter, we work with an iterative setting for a coherent quantum memory under limited quantum control; more specifically, we propose a scheme to sequentially transfer the states between a number of flying qubits and a long (memory) chain of $N$ identical, ferromagnetically polarised, but non-interacting, static spins in a {\it passive} way (see Fig.~\ref{fig:device}). 

We start by considering the first flying qubit in some arbitrary state $\ket{\psi_1}_{f_1} = \alpha_1 \ket{\uparrow} + \beta_1 \ket{\downarrow}$ and its associated QST process with the static spins in the ferromagnetic chain $\ket{F}_c = \ket{\uparrow_{s_1} ... \uparrow_{s_N}}$. As shown in Fig.~\ref{fig:device}, during encoding (or the `write' operation) $\ket{\psi_1}_{f_1}$ enters from the front into the chain and interacts with the static spins $s_1$ to $s_N$ sequentially. We model the interactions by an effective Hamiltonian coupling each flying qubit and the $k^{th}$ static spin of the following form: 
\begin{equation}
H_{k} = \frac{g_{k}}{2} (\sx^f \sx^{s_{k}} + \sy^f \sy^{s_{k}}) = g_{k} (\sp^f \sm^{s_{k}} + \sm^f \sp^{s_{k}}),
\label{eq:hamil}
\end{equation}
where $\sigma_{\pm} = (\sx \pm i\sy)/2$ are Pauli spin-flip operators. The $g_{k}$ are the $XY$ exchange coupling strengths that depend on the separation of the qubit and the spin $s_k$, and are hence time dependent due to the mobile qubit. Thus for a general state $\ket{\Psi_{k} (t)} = U_{k} (t) \ket{\Psi_{k} (0)}$ of the qubit and the spin $s_k$, the time evolution operator $U_{k} (t) = \exp\left[- i \theta_k (t) (\sp^f \sm^{s_{k}} + \sm^f \sp^{s_{k}})\right]$, where $\theta_{k} (t) = \int_{0}^{t} g_{k}(t') dt'/\hbar$ is constant when $[0, t]$ is chosen so that $g_{k} (0)$ and $g_{k} (t)$ are negligible. In the basis $\ket{\uparrow_f \uparrow_{s_{k}}}$, $\ket{\uparrow_f \downarrow_{s_{k}}}$, $\ket{\downarrow_f \uparrow_{s_{k}}}$, $\ket{\downarrow_f \downarrow_{s_{k}}}$,
\begin{equation}
U_{k} =
\left( {\begin{array}{cccc}
1 & 0 & 0 & 0 \\
0 & \cos \theta_{k} & - i \sin \theta_{k} & 0 \\
0 & - i \sin \theta_{k} & \cos \theta_{k} & 0 \\
0 & 0 & 0 & 1 \\
\end{array}}\right).
\label{eq:unitary}
\end{equation}
For the {\it Heisenberg} model with an extra term $\frac{g_k}{2} \sigma_z^f \sigma_z^{s_k}$ in Eq.~\ref{eq:hamil}, $U_k$ takes the same form apart from an extra phase factor of $e^{i \theta_k}$ for each trigonometric term. Let us first consider the $XY$ case with $\theta_k = \theta \in (0, \pi/2]$ $\forall\ k$. 

Starting from the total state $\ket{\psi_1, F} = \ket{\psi_1}_{f_1} \ket{F}_c$, we apply $U_k$ as in Eq.~\ref{eq:unitary} to the flying qubit and the $k^{th}$ static spin sequentially for $s_1$ through to $s_N$. The state $\ket{\uparrow, F}$ with amplitude $\alpha_1$ remains the same during this write operation, while $\ket{\downarrow, F}$, with amplitude $\beta_1$, evolves as follows:
\begin{eqnarray}
\hspace{-4mm}  &\ket{\downarrow, F}& \Rightarrow_1  \cos \theta\ \underline{\ket{\downarrow, F}} - i \sin \theta\ \widehat{S}_{1}^{-} \ket{\uparrow, F} \nonumber \\
\hspace{-5mm}  & \Rightarrow_2 & \cos \theta \big(\cos \theta\ \underline{\ket{\downarrow, F}} - i \sin \theta\ \widehat{S}_{2}^{-} \ket{\uparrow, F} \big) - i \sin \theta\ \widehat{S}_{1}^{-} \ket{\uparrow, F} \nonumber \\
\hspace{-5mm}  & \Rightarrow_k & ... \ \forall \ k = 3, ... , N-1 \nonumber \\
\hspace{-5mm}  & \Rightarrow_N & - i \sin \theta \sum_{k=1}^{N} \cos^{k-1} \theta \ \widehat{S}_{k}^{-} \ket{\uparrow, F} + \cancelto{0\ as\ N \rightarrow \infty}{\cos^{N}} \theta \ket{\downarrow, F}
\label{eq:der}
\end{eqnarray}
where $\Rightarrow_k$ corresponds to the occurrence of an interaction event for the flying qubit with the static spin $s_k$, and $\widehat{S}_{k}^{-}$ denotes a spin down-flip in the $k^{th}$ position of the chain. Note that the only component which evolves further in each superposition is underlined in Eq.~\ref{eq:der}. Combining both parts, we see that the flying qubit emerges as polarised $\ket{\uparrow}_{f_1}$ with probability $1 - \cos^{2N} \theta \rightarrow 1$ for $N \rightarrow \infty$. In this limit, the initial quantum information which $\ket{\psi_1}_{f_1}$ held before encoding has now been transferred to the chain whose collective state reads
\begin{equation}
\bigg( \alpha_1 - i \beta_1 \sin \theta \sum_{k=1}^{N \rightarrow \infty} \cos^{k-1} \theta \ \widehat{S}_{k}^{-} \bigg) \ket{F}.
\label{eq:c1}
\end{equation}
This is model independent by conservation of total spin, and so the argument is equally valid for the Heisenberg model (up to factors of $e^{i k \theta}$). Note that the probability of a single down-flip at site $k$ is $|\beta_1|^2 \sin^2 \theta \cos^{2(k-1)} \theta$, a quantity that decays exponentially along the chain. Summing over all these probabilities gives $|\beta_1|^2$, as expected by conservation of total probability.

To read out the state $\ket{\psi_1}$ of the original flying qubit from the memory at a later stage, we inject a polarised flying qubit $\ket{\uparrow}_f$ from the tail back to the chain, i.e., in the opposite direction as for the encoding operation (see Fig.~\ref{fig:device}). This is the simplest decoding method for the memory system, especially when the level of quantum control is limited. The flying qubit sequentially interacts with the static spins through $s_N$ to $s_1$ and, provided it has the same kinetic energy as the emitted qubit $f_1$, the total state of the system after decoding becomes
\begin{eqnarray}
\hspace{-7mm} \bigg(\alpha_1 \ket{\uparrow, F} - \beta_1 \underbrace{\sin^2 \theta \sum_{k=1}^{\infty} \cos^{2 (k-1)} \theta}_1 \ket{\downarrow, F} \bigg) - i \beta_1 \sin \theta \times \nonumber \\
\hspace{-7mm} \cancelto{0}{\bigg( \sum_{k=1}^{\infty}  \cos^{k} \theta \ \widehat{S}_{k}^- - \sin^2 \theta \sum_{n=2}^{\infty} \sum_{j=1}^{n-1} \cos^{n+j-2} \theta \ \widehat{S}_{n-j}^- \bigg)} \ket{\uparrow, F}.\
\label{eq:read}
\end{eqnarray}
This is straightforwardly proved by applying the two-spin unitary operations sequentially as detailed in the Supplementary Material~\cite{sm}.

Thus, after the read operation the chain returns to the original ferromagnetic state and is disentangled from the flying qubit, which now emerges from the front as
\begin{equation}
\sz^{f} \ket{\psi_1}_{f_1} = \alpha_1 \ket{\uparrow} - \beta_1 \ket{\downarrow}.
\label{eq:phase}
\end{equation}
In other words we have recovered the original state of this flying qubit, up to a phase flip which can be corrected by a simple $\sz^{f}$ gate. This will always be the case for the $XY$ model, and importantly distinguishes the decoding process from a time reversal operation. However, the same read operation does not work for a Heisenberg-type coupling (except for special cases), since the extra phases $e^{i k' \theta}$ invalidate the cancellation in Eq.~\ref{eq:read}.

\vspace{0.5mm}
To see how the size $N$ of the memory chain required to store one qubit scales with the coupling strength $\theta$, we note that the only condition is $\cos^{2N} \theta \rightarrow 0$ for sufficiently large $N$. Given any error tolerance $\epsilon > 0$ such that $\cos^{2N} \theta < \epsilon$,  we require
\begin{equation}
2 N (\theta) > \frac{\ln \epsilon}{\ln \cos \theta}\approx -\frac{2\ln \epsilon}{\theta^2} \hspace{2mm} (\theta\ll1).
\label{eq:tolerance}
\end{equation}
Thus we see that the required memory size increases significantly with decreasing coupling strength $\theta$ for fixed $\epsilon$. For example, with $\epsilon \sim 10^{-4}$ and $\theta=1$, $N_{\text{min}} \simeq 8$ rising to $\sim 920$ for $\theta=0.1$. However, in the weak coupling regime, the qubit state stored in the chain is delocalised; each static spin only shares a fraction of the total information, and thus has the potential to store more. On the other hand, for strong couplings a relatively short chain can already store a number of flying qubits. In the special case where $\theta = \frac{\pi}{2}$, one polarised static spin in the chain is sufficient to store a flying qubit, since now $U_k$ in Eq.~\ref{eq:unitary} is simply a {\it SWAP}$_k$ gate (up to a phase of `$- i$'). Encoding of subsequent flying qubits is thus also obvious, as are their readout procedures (since {\it SWAP}$_k^2 =$ I up to a phase of `-1'). These qubits are stored locally in the memory: By induction and linearity, one finds that $n$ ordered flying qubits $\ket{\psi_n}_{f_n} ... \ket{\psi_1}_{f_1}$ are encoded sequentially into the memory and the chain state becomes $\ket{\tilde{\psi}_n}_{s_1} ... \ket{\tilde{\psi}_1}_{s_n} \ket{\uparrow ... \uparrow}_s$ after the whole write operation. Once read back, we obtain $\ket{\tilde{\psi}_n}_{f_n} ... \ket{\tilde{\psi}_1}_{f_1}$, where $\tilde{ }$ denote the appropriate phase changes for the $XY$ model. For this special case the argument also holds for the Heisenberg exchange model, though the extra phase flips will not be present. Indeed in the Heisenberg case the decoding process is nothing but a time reversal of the encoding operation. In fact, realisation of time reversal ($\theta \rightarrow - \theta$) for arbitrary $\theta$ would make the memory system viable even for the Heisenberg case in general. However, this would require tuning from ferromagnetic coupling $g_k$ (write) to anti-ferromagnetic $- g_k$ (read), which is impractical.

\begin{figure}[h]
\begin{center}$
\begin{array}{cc}
  \hspace{-4.8mm} \subfigure[\ $N_s = 1$]{\includegraphics[width=1.61in]{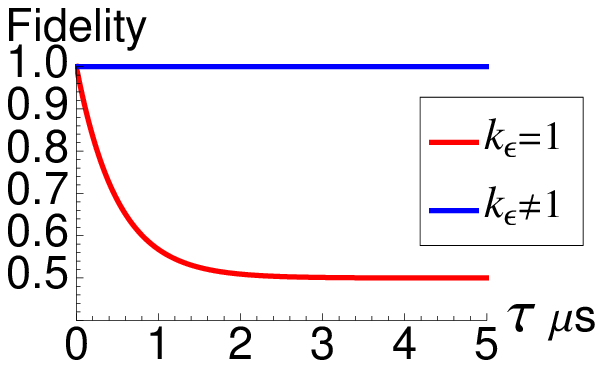}} &
  \hspace{-1.2mm} \subfigure[\ $N_s = 100$]{\includegraphics[width=2.03in]{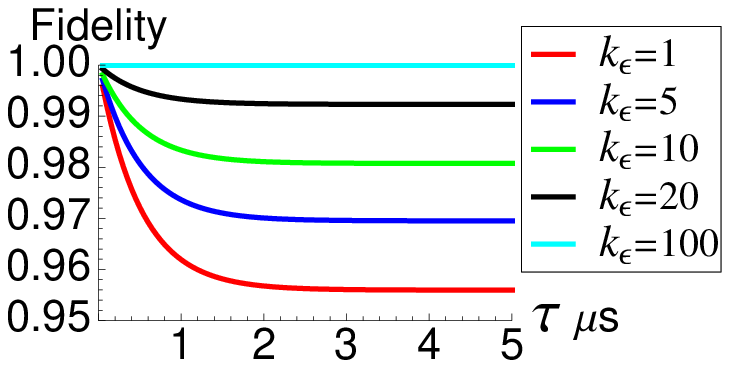}}  
\end{array}$
\caption{Plots for the fidelity of the retrieved qubit, relative to the input $\ket{+}_{f_1}$, against the storage time $\tau$ under inhomogeneous dephasing with rate $\Gamma = 1$~MHz (coherence time 1$~\mu$s) for the $k_{\epsilon}^{th}$ static spin and {\it zero} for all others ($N=100$). We vary $\theta$ such that the qubit is stored in the first $N_s$ number of static spins with $\epsilon \sim 10^{-2}$ (see Eq.~\ref{eq:tolerance}): (a) The qubit is stored entirely in the first qubit ($\theta = \pi/2$); (b) The whole chain of 100 spins store the qubit collectively ($\theta \simeq 0.30$).}
\label{fig:dephasing}
\end{center}
\end{figure}

\vspace{0.5mm}
We now consider the effect of decoherence on a chain in which one qubit state $\ket{+}_{f_1} = \big(\ket{\uparrow}_{f_1} + \ket{\downarrow}_{f_1}\big)/\sqrt{2}$ is stored. Both the write and the read operations are assumed to be fast; the principal effect of decoherence is on the static spins for the time ($\tau$) during which the qubit is stored in the chain. We model this using the Lindblad master equation~\cite{breuer02}, and by restricting to the subspace consisting of only zero or one excitations among the static spins. We simulated the behaviour for the memory chain of a fixed length $N = 100$ under dephasing errors. We also vary $\theta$ such that the qubit is stored in the first $N_s$ ($\leq N$) number of static spins with $\epsilon \sim 10^{-2}$ (see Eq.~\ref{eq:tolerance}), to monitor the effects of spreading the quantum information stored in the memory. For homogeneous dephasing, in which each spin is subject to the same, independent decoherence process, the total decoherence for the memory is essentially the same for all $\theta$ values. In our simulations we take a dephasing rate $\Gamma = 1$~MHz for each static spin and each resulting curve coincides exactly with the red curve in Fig.~\ref{fig:dephasing}(a), regardless of the chosen $N_s\leq N$. In other words, the decoherence rate does not depend on how local or distributed the information is in the quantum memory, since the relevant qubits which contain the quantum information each decohere at the same rate $\Gamma$. It also follows that the fidelity must eventually saturate to $\frac{1}{2}$ when all the memory qubits lose their quantum information.

On the other hand, for inhomogeneous  decoherence processes in which the spins decohere at different rates, distributing the quantum information does reduce the variance of the total memory decoherence rate compared with locally stored information, and this difference can be very large.  We illustrate this in Fig.~\ref{fig:dephasing} for the case when just one static spin is subject to decoherence (for example due to the proximity of a magnetic impurity). When the quantum information is stored locally (in the first qubit with $\theta=\frac{\pi}{2}$ in Fig.~\ref{fig:dephasing}(a)) then the fidelity $F$ saturates to $\frac{1}{2}$ when the decohering spin is the first, whereas $F = 1$ if the decohering spin is any of the other $N -$ 1 spins. However, when the quantum information is spread over all $N$ spins ($\theta \simeq 0.30$ in Fig.~\ref{fig:dephasing}(b)), the saturated fidelity becomes  $\gtrsim 1 - \frac{1}{2N}$, independent of which spin is subject to the decoherence.  This result may easily be extended to cases when more than one spin decoheres, or the spins decohere at different rates. Spreading the information over many qubits will again {\it smooth} the statistical fluctuations in information loss.

\vspace{1mm}
The general result for encoding and decoding an arbitrary number of input flying qubits with arbitrary multi-particle superpositions remains presently a conjecture, but we have strong evidence that it is true. In particular, we have extended the proof from one to two flying qubits, including entangled states, as well as for the case where there is only one $\ket{\downarrow}_f$ among any number of flying $\ket{\uparrow}_f$ qubits. Both of these proofs can be found in the Supplementary Material~\cite{sm}, but we outline some essential ingredients of them here.

For the second of the two proofs just mentioned, we define a 0$^{th}$ collective 1-spin down-flip operator on the chain $\ket{F}_c$ arising from Eq.~\ref{eq:der},
\begin{equation}
\widehat{D}^{(1)}_0 = \sum_{k=1}^{\infty} a^{(1)}_0(k)\ \widehat{S}_{k}^{-} := \sum_{k=1}^{\infty} (- i \sin \theta \cos^{k-1} \theta)\ \widehat{S}_{k}^{-}.
\end{equation}
We find $\sum_k |a^{(1)}_0(k)|^2 = 1$, as expected for unit total probability. We then show that this chain distribution can be altered by further storing $l$ subsequent $\ket{\uparrow}_f$ qubits, resulting in the $l^{th}$ collective 1-spin down-flip,
\begin{equation}
\widehat{D}^{(1)}_{l} = \sum_{k=1}^{\infty} a^{(1)}_{l}(k)\ \widehat{S}_{k}^{-}
\end{equation}
with
\begin{eqnarray}
\hspace{-1mm} a^{(1)}_{l}(k) &=& a^{(1)}_0(k) \sum_{r=0}^{\text{min}\{l, k-1\}} (-1)^r \binom{k-1}{r} \binom{l}{r} \tan^{2r} \theta\ \cos^{l} \theta \nonumber \\
&=& a^{(1)}_0(k)\ _2F_1 (1-k, -l; 1; - \tan^2 \theta)\ \cos^{l} \theta
\label{eq:coef}
\end{eqnarray}
where the more compact form $_2F_1 (a, b; c; z)$ denotes the {\it Gauss} hypergeometric function~\cite{beals10}. The detailed derivation is given in the Supplementary Material, together with the proof for decoding the stored qubits in the reverse order~\cite{sm}. This also means the ferromagnetic chain can store the two-qubit states $\ket{\uparrow_{f_1} \downarrow_{f_2}}$, $\ket{\downarrow_{f_1} \uparrow_{f_2}}$, which can be read out sequentially. 

\begin{figure}[h]
\begin{center}$
\begin{array}{cc}
 \hspace{-2mm} \subfigure[\ $1 - \text{Fidelity} = 3 \times 10^{-4}$, $\ket{\Phi^-}_{f_{12}} = (\ket{\uparrow \uparrow} - \ket{\downarrow \downarrow})/\sqrt{2}$;]{\includegraphics[width=1.5in]{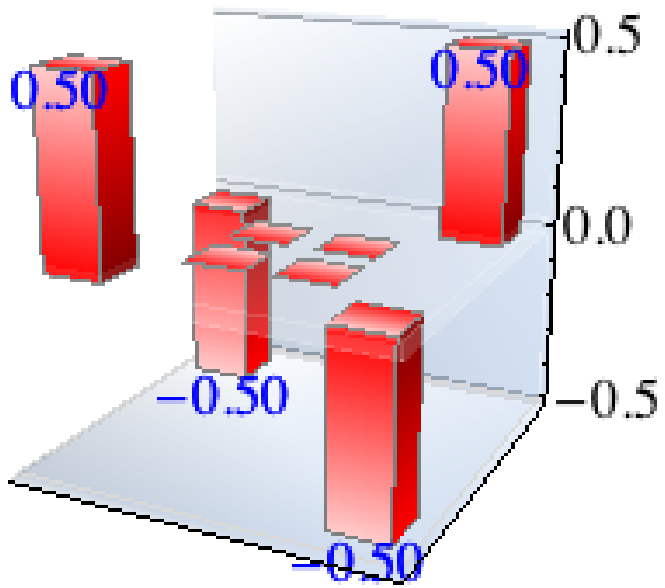}} &
 \hspace{2mm} \subfigure[\ $1 - \text{Fidelity} = 7 \times 10^{-4}$, $\ket{\Psi^-}_{f_{12}} = (\ket{\uparrow \downarrow} - \ket{\downarrow \uparrow})/\sqrt{2}$.]{\includegraphics[width=1.42in]{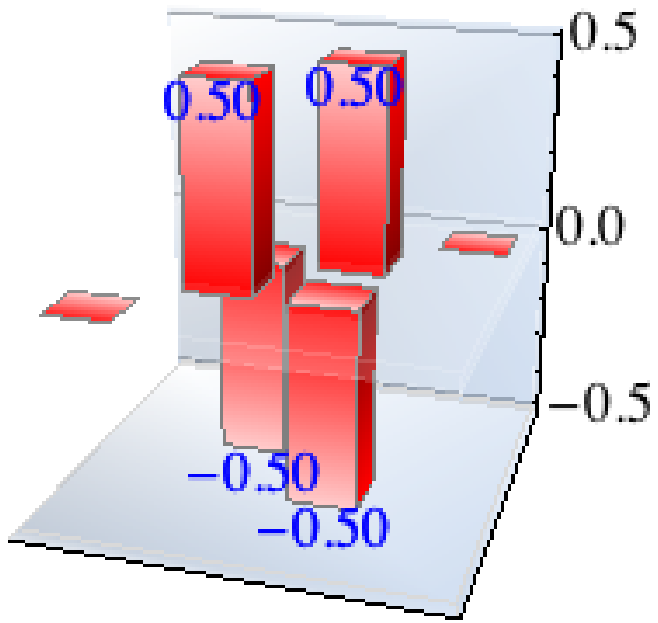}}
\end{array}$
\caption{Density state tomograms constructed for the corresponding, retrieved (two-qubit) quantum state from a memory chain $\ket{F}_c$ consisting of $N =8$ static spins without decoherence. (a) and (b) represent two separate cases of memory storage and retrieval of quantum information. The inputs, both entangled, are indicated in each case, together with the corresponding infidelity of the retrieved state. We chose a non-special value $\theta = 1.1$; for a tolerance $\epsilon \sim 10^{-4}$, this requires $N \geq 6$ to reliably store one qubit, as predicted by Eq.~\ref{eq:tolerance}.}
\label{fig:sim1}
\end{center}
\end{figure}

We have also proved that encoding and decoding two $\ket{\downarrow}_{f}$ qubits in the chain $\ket{F}_c$ can be done in a similar fashion~\cite{sm}, with the $(0, 0)^{th}$ collective 2-spin down-flip amplitude \big(after encoding only $\ket{\downarrow_{f_1} \downarrow_{f_2}}$\big) being
\begin{equation}
\hspace{-3mm} a^{(2)}_{(0,0)} (k_1,k_2) = (-i \sin \theta)^2 \cos^{k_1+k_2-1} \theta\ (2-(k_2 - k_1 - 2) \tan^2 \theta).
\end{equation}
By linearity, the memory can thus store at least two qubits of arbitrary states, including any bipartite entanglement (and hence qubits of mixed states). A numerical check for $N=8$ static spins and two fully entangled flying qubits is shown in Fig.~\ref{fig:sim1}, demonstrating that this state can be retrieved with high fidelity and errors within the expected limits.  

To generalize the derivations, we need to show that the chain can store a spin state that includes an arbitrary number $n$ of down-spins
\begin{equation}
\hspace{-.5mm} \widehat{D}^{(n)}_{(l_1, ..., l_n)} = \sum_{k_1 < ... < k_n} a^{(n)}_{(l_1, ..., l_n)} (k_1, ..., k_n)\ \widehat{S}^-_{(k_1, ..., k_n)}
\end{equation}
where the $k_i$s denote the spin-flip positions and $l_i$ denotes the number of $\ket{\uparrow}_f$ encoded between the $i^{th}$ and $(i+1)^{th}$ flying $\ket{\downarrow}_f$ qubits. This would mean that the chain can store a number of flying qubits, each of which was originally either $\ket{\uparrow}_f$ or $\ket{\downarrow}_f$ and can be further retrieved by the aforementioned decoding mechanism. Arbitrary multi-particle superpositions and entanglement then follows by linearity. Unfortunately, this procedure becomes impractical for more than 2 down-spins due to the difficulty of keeping track of all the indices, and a rigorous proof for an arbitrary number of up-spin and down-spin qubits remains an open challenge. 

\begin{figure}[h]
\begin{center}
\includegraphics[width=2in]{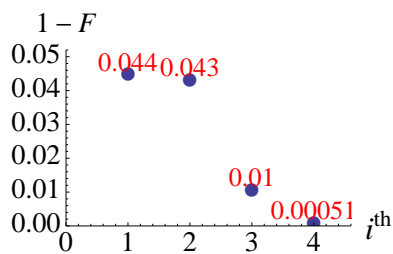}
\caption{Plot for the infidelity $1 - F$ of the retrieved qubit (with phase corrected), relative to the corresponding input $\rho_i$, against the ordinal number $i$ of the inputs. Here, four randomly-generated pure qubits \{$\rho_i$\}$_4$ are sequentially encoded into the memory chain $\ket{F}_c$ ($N=9$), and then retrieved one by one in the reverse order without decoherence. We used $\theta = 1$, and hence for $\epsilon \sim 10^{-4}$ (\text{or} $10^{-2}$) it requires $N \geq 7$ (or 4) to reliably store one qubit. 
The finite chain length restricts the number of qubits the memory can hold; within its capacity the last encoded (or the first retrieved) qubit always has the highest fidelity, since part of the quantum information stored for the earliest qubits may have been ``pushed out'' of the memory by the later ones during encoding. 
}
\label{fig:sim2}
\end{center}
\end{figure}

However, further support for the conjecture is provided by numerical simulations as shown in Fig.~\ref{fig:sim2} for short chains with a few flying qubits of randomly generated states, and $\theta$ chosen to give a high probability for a ferromagnetic tail after encoding. Analysis of the fidelity after decoding gives results which are consistent with the conjecture within the expected error bounds. 

\begin{figure}[h]
\begin{center}
\includegraphics[width=1.6in]{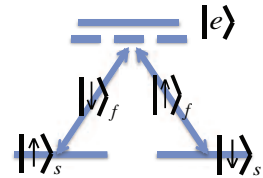}
\caption{The static pseudo-spin structure, with a ground doublet plus a virtual, excited level: The XY exchange model between the flying photon and each QD can thus be achieved.}
\label{fig:raman}
\end{center}
\end{figure}

\vspace{1mm}
We conclude with a discussion of a potential architecture for the memory. In essence, it is a slight extension of the cavity quantum electrodynamics system suggested in~\cite{ciccarello08}. The qubits correspond to the polarization states of photons, which are transported reliably in a semiconductor photonic waveguide. An array of embedded quantum dots (QDs) act as the chain, where each multi-level QD exhibits an effective pseudo spin-1/2 within its degenerate doublet ground state; an $XY$ exchange model is achieved via two-photon Raman processes involving virtually populating an excited level (see Fig.~\ref{fig:raman})~\cite{ciccarello08}. A low operating temperature on the mK scale is required to initialise the memory, and also to have a long coherence time ($\sim$$\mu$s) for each QD and hence the memory. With weak coupling between the photon and the QDs, the stored qubits are delocalised in the memory: the QDs store the qubits collectively. The QDs should also be well separated to have negligible interactions with each other, say spaced by 10 to 100~nm. In that case, within one millimetre there can be as many as $10^4$$\sim$$10^5$ non-interacting QDs in the memory, where it only takes 10~ps for each write/read operation. In this optical structure, $\theta$ variations are also negligible, as are the variations in the kinetic energy of the photons. Therefore, this memory structure is resilient to various imperfections. It should passively store qubits much more quickly, and have less photon loss, than previous holographic schemes~\cite{wesenberg09} since no active control is needed during successive write/read operation rounds. Other systems such as optical lattices, where $XY$ exchange interactions can be effectively achieved (and tuned), with an internal electronic degree of freedom of the cold atoms serving as both the static spins and the flying qubits~\cite{duan03}, are also potential candidates for the memory realization.

\vspace{0.5mm}
\begin{acknowledgements}
B.W.L. acknowledges the Royal Society for a University Research Fellowship. Y.P. thanks Hertford College, Oxford for a scholarship.
\end{acknowledgements}


\end{document}


\author{Yuting Ping}
\email{yuting.ping@materials.ox.ac.uk}
\affiliation{Department of Materials, University of Oxford, Oxford OX1 3PH, United Kingdom}
\author{John H. Jefferson}
\affiliation{Department of Physics, Lancaster University, Lancaster LA1 4YB, United Kingdom}
\author{Brendon W. Lovett}
\email{b.lovett@hw.ac.uk}
\affiliation{School of Engineering and Physical Sciences, Heriot-Watt University, Edinburgh EH14 4AS, United Kingdom}
\affiliation{Department of Materials, University of Oxford, Oxford OX1 3PH, United Kingdom}

\title{SUPPLEMENTARY MATERIAL\\ A coherent and passive one dimensional quantum memory}
\maketitle

\section{Read-out of the first qubit}
In the main text, we have shown that the quantum information held by the first flying qubit $\ket{\psi_1}_{f_1}$ can be encoded into the memory chain via the proposed mechanism. The collective state of the chain after encoding is described by Eq.~4 in the main text, while the flying qubit emerges as polarised $\ket{\uparrow}_{f_1}$. We further proposed that the state $\ket{\psi_1}$ can be decoded back from the chain to a polarised flying qubit $\ket{\uparrow}_f$ injected with the same kinetic energy in the `read' direction (see Fig.~1 in the main text). Here, we shall derive the total state of the system after decoding (Eq.~5 in the main text), which provides further insight into the nature of the spin chain memory. 

Just before the read operation, the total state of the flying qubit and the chain is
\begin{equation}
\bigg( \alpha_1 - i \beta_1 \sin \theta \sum_{k=1}^{N \rightarrow \infty} \cos^{k-1} \theta \ \widehat{S}_{k}^{-} \bigg) \ket{\uparrow, F}.
\label{eq:befread1}
\end{equation}
The state $\ket{\uparrow, F}$ with amplitude $\alpha_1$ again remains the same during decoding, while each state $\widehat{S}_{k}^{-} \ket{\uparrow, F}$, with amplitude $-i \beta_1 \sin \theta \cos^{k-1} \theta$, evolves as follows, as the flying spin passes each member of the chain:
\begin{align}
&\widehat{S}_{k}^{-} \ket{\uparrow, F} \Longrightarrow_{N (\rightarrow \infty)} ... \Longrightarrow_{k+1} \widehat{S}_{k}^{-} \ket{\uparrow, F} \nonumber \\ 
& \Longrightarrow_{k} \hspace{3.5mm} \cos \theta \ \widehat{S}_{k}^{-} \ket{\uparrow, F} - i \sin \theta\ \underline{\ket{\downarrow, F}} \nonumber \\
& \Longrightarrow_{k-1}  \cos \theta \ \widehat{S}_{k}^{-} \ket{\uparrow, F} - i \sin \theta\ \times \nonumber \\
& \hspace{36mm} \big(\cos \theta\ \underline{\ket{\downarrow, F}}  - i \sin \theta\ \widehat{S}_{k-1}^{-} \ket{\uparrow, F} \big) \nonumber \\
& \Longrightarrow_{k'}  ... \hspace{1mm} \forall\ k' = k-2, ..., 2 \nonumber \\
& \Longrightarrow_{1} \hspace{3.5mm} \cos \theta \ \widehat{S}_{k}^{-} \ket{\uparrow, F} - i \sin \theta \cos^{k-1} \theta \ket{\downarrow, F}  \nonumber \\
& \hspace{8.5mm} - \sin^{2} \theta \sum_{j=1}^{k-1} \cos^{j-1} \theta \ \widehat{S}_{k-j}^{-} \ket{\uparrow, F} \hspace{4mm} \text{if $k \geq 2$} \nonumber \\
& \hspace{2mm} \text{or} \hspace{7mm} \cos \theta \ \widehat{S}_{1}^{-} \ket{\uparrow, F} - i \sin \theta \ket{\downarrow, F} \hspace{5mm} \text{if $k = 1$}
\label{eq:read1}
\end{align}
where we have followed the same notation as in the main text. Thus, by linearity, the total state after decoding becomes 
\begin{align}
& \hspace{-2mm} \alpha_1 \ket{\uparrow, F} - i \beta_1 \sin \theta\ \big(\cos \theta \ \widehat{S}_{1}^- \ket{\uparrow, F} - i \sin \theta \ket{\downarrow, F} \big) \nonumber \\
& \hspace{-7.5mm} - i \beta_1 \sin \theta \sum_{k=2}^{\infty} \cos^{k-1} \theta \big(\cos \theta \ \widehat{S}_{k}^- \ket{\uparrow, F} - i \sin \theta \cos^{k-1} \theta \ket{\downarrow, F} \big) \nonumber \\
& \hspace{-7.5mm} + i \beta_1 \sin \theta \sum_{n=2}^{\infty} \cos^{n-1} \theta \big( \sin^2 \theta \sum_{j=1}^{n-1} \cos^{j-1} \theta \ \widehat{S}_{n-j}^- \ket{\uparrow, F}\big)
\label{eq:afread1}
\end{align}
where in the last line we have replaced the dummy variable $k$ by $n$. By absorbing the bracketed terms in the first line into the summation of the second line, Eq.~\ref{eq:afread1} is equivalent to
\begin{align}
& \bigg(\alpha_1 \ket{\uparrow, F} - \beta_1 \underbrace{\sin^2 \theta \sum_{k=1}^{\infty} \cos^{2 (k-1)} \theta}_1 \ket{\downarrow, F} \bigg) - i \beta_1 \sin \theta\ \times \nonumber \\
& \bigg( \sum_{k=1}^{\infty}  \cos^{k} \theta \ \widehat{S}_{k}^- - \sin^2 \theta \sum_{n=2}^{\infty} \sum_{j=1}^{n-1} \cos^{n+j-2} \theta \ \widehat{S}_{n-j}^- \bigg) \ket{\uparrow, F},
\label{eq:afread1a}
\end{align}
(i.e., Eq.~5 in the main text) where the first sum is simply a geometric series. For the double summation we want to focus on the  coefficients for $\widehat{S}_{k}^-$, and hence the sum over $j$ can be replaced by a sum over $k$ with $k=n-j \geq 1$, i.e.,  
\begin{align}
\sum_{n=2}^{\infty} \sum_{j=1}^{n-1} \cos^{n+j-2} \theta \ \widehat{S}_{n-j}^- & = \sum_{k=1}^{\infty} \sum_{n=k+1}^{\infty} \cos^{2(n-1)-k} \theta \ \widehat{S}_{k}^- \nonumber \\
& = \sum_{k=1}^{\infty} \frac{\cos^k \theta}{\underbrace{1-\cos^2 \theta}_{\sin^2 \theta}} \ \widehat{S}_{k}^-
\end{align}
where we have swapped the double summations as the sum over $n$ runs to infinity. Therefore, the second line in Eq.~\ref{eq:afread1a} becomes zero due to complete cancellations for each $k=1, 2, ...\ \infty$, and the total state after decoding is then simply
\begin{equation}
\big(\alpha_1 \ket{\uparrow} - \beta_1 \ket{\downarrow}\big)_f \bigotimes \ket{F}_c.
\end{equation}

The complete cancellations only require that $N \rightarrow \infty$, which is the same condition as for the encoding procedure. Thus in this limit, the state of the flying qubit with matching kinetic energy can be transferred to and back from the memory in the way we have proposed in the main text. Note that this read operation is not a time reversal in general.

If we employ the Heisenberg model, the state of the flying qubit can still be encoded into the chain, as explained in the main text. However, once read back, the total state of the system becomes (by tracking the extra phases $e^{i k' \theta}$ in the above derivation),
\begin{equation}
\big(\alpha_{1} \ket{\uparrow} + \beta'_{1} \ket{\downarrow} \big)_f \bigotimes \ket{F}_c + \sum_{k=1}^{\infty} \gamma_{k} \ket{\uparrow}_f \bigotimes \ \widehat{S}_{k}^- \ket{F}_c
\label{eq:afreadheisen}
\end{equation}
where
\begin{equation*}
\beta'_{1} = - \beta_1 \sin^2 \theta \sum_{k=1}^{\infty} e^{2 i k \theta} \cos^{2 (k-1)} \theta = \frac{- \beta_1 e^{2 i \theta} \sin^2 \theta}{1 - e^{2 i \theta} \cos^2 \theta},
\end{equation*}
\begin{align}
\gamma_k & =  - i \beta_1 e^{i \theta} \sin \theta\ \bigg( e^{i k \theta} \cos^k \theta\ -  \nonumber \\
& \hspace{27mm} \sin^2 \theta \sum_{n=k+1}^{\infty} e^{i (2n-k) \theta} \cos^{2(n-1)-k} \theta \bigg) \nonumber \\
& = - i \beta_1 \sin \theta \cos^k \theta\  e^{i (k+1) \theta}  \bigg(\frac{1- e^{2 i \theta}}{1 - e^{2 i \theta} \cos^2 \theta} \bigg) \tag{S7a}
\end{align}
for all $k \in \mathbb{Z}^+$. The decoding procedure no longer works due to the extra phases, except when $\theta = \frac{\pi}{2}$. Note that Eq.~\ref{eq:afreadheisen} is consistent with the $XY$ model once we identify each extra phase $e^{i \theta}$ with 1 accordingly.

\begin{figure}[h]
\begin{center}
\includegraphics[width=3.3in]{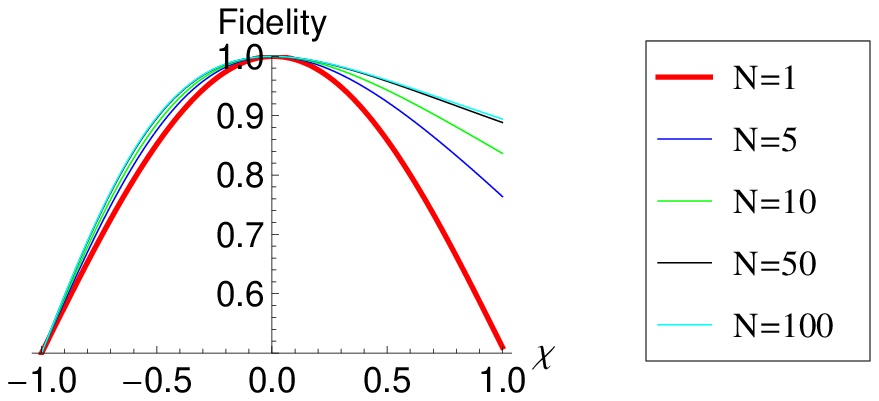}
\caption{Plots of the fidelity of the retrieved qubit, relative to the input $\ket{+}_{f_1}$, against the fractional difference $\chi$ between the coupling strengths during encoding ($\theta_{enc}$) and decoding ($\theta_{dec}$), for various chain lengths. For each $N$, the whole chain collectively stores the qubit, i.e., $N_s = N$.}
\label{fig:chi}
\end{center}
\end{figure}

\vspace{1mm}
When the kinetic energy of the flying qubit during the read operation differs from that of the input for encoding (i.e., $\theta_{dec} \neq \theta_{enc}$), we numerically simulate how the error effects on the memory for various chain lengths $N$ in Fig.~\ref{fig:chi}, where $\chi = (\theta_{dec} - \theta_{enc})/\theta_{enc}$. Here, $\theta_{enc}$ is adjusted so that each chain with length $N$ is just enough to collectively store the qubit (it is defined by setting $\cos^N \theta_{enc}=0.01$).

From Fig.~\ref{fig:chi}, we see that after the decoding round, the retrieved qubit is of a high fidelity ($\geq 99$\%) with respect to the original input, for small mismatches ($\sim$10\%) in the encoding and decoding $\theta$ values. Moreover, as the number $N$ of static spins increases, the memory's tolerance to such errors improves.

\section{Collective 1-spin down-flip distributions}
In the main text, we introduced the $l^{th}$ collective 1-spin down-flip distribution $\widehat{D}_l^{(1)} = \sum_k a_l^{(1)} (k) \widehat{S}_k^-$ in Eqs.~8-10; $|a_l^{(1)} (k)|^2$ corresponds to the probability of the $k^{th}$ static spin being $\ket{\downarrow}_{s_k}$ in the respective distribution. 

Arising from the chain state after encoding the first flying $\ket{\downarrow}_{f_1}$ qubit (see Eq.~3 in the main text), the $0^{th}$ 1-spin down-flip distribution has $a_0^{(1)} (k) = -i \sin \theta \cos^{k-1} \theta$. The total probability is $\sum_{k=1}^{\infty} |a_0^{(1)} (k)|^2 = 1$, while the mean position $\mu^{(1)}_0$ of the down-flip and the associated standard deviation $\sigma^{(1)}_0$ are (see Appendix I)
\begin{align}
&\mu^{(1)}_0 = \sum_{k=1}^{\infty} k\ |a^{(1)}_0(k)|^2 = \csc^2 \theta \nonumber \\
&\sigma^{(1)}_0 = \sqrt{\sum_{k=1}^{\infty} k^2 |a_0^{(1)}(k)|^2 - (\mu^{(1)}_0)^2} = \cos \theta \csc^2 \theta. 
\label{eq:mean}
\end{align}

The $l^{th}$ collective 1-spin down-flip distribution in the memory results from further encoding $l$ subsequent flying $\ket{\uparrow}_f$ qubits. We now inductively derive the analytical expression for $a^{(1)}_l (k)$ given by Eq.~10 in the main text.

\vspace{0.5mm}
First, we find that after encoding one flying $\ket{\uparrow}_f$ qubit into a chain with state $\widehat{S}_k^- \ket{F}_c$ $\forall$ finite $k \ll N$, the total state becomes
\begin{equation}
\ket{\uparrow}_f \bigotimes \big(\cos \theta\ \widehat{S}_k^- - \sin^2 \theta \sum_{n=k+1}^{\infty} \cos^{n-k-1} \theta\ \widehat{S}_n^- \big) \ket{F}_c 
\label{eq:wrop1}
\end{equation}
which is derived step-by-step as before. Thus, by linearity (apply Eq.~\ref{eq:wrop1}), encoding one flying $\ket{\uparrow}_f$ qubit into the chain with distribution $\widehat{D}_0^{(1)} = \sum_k a_0^{(1)} (k) \widehat{S}_k^-$ gives rise to the first 1-spin down-flip distribution
\begin{equation}
\hspace{-.5mm} \widehat{D}_1^{(1)}  = \sum_{k=1}^{\infty} a_0^{(1)} (k) \big(\cos \theta\ \widehat{S}_k^- - \sin^2 \theta \sum_{n=k+1}^{\infty} \cos^{n-k-1} \theta\ \widehat{S}_n^- \big).
\label{eq:l1}
\end{equation}
Again, we want to focus on the coefficients of $\widehat{S}_k^-$ terms for the double sum. Note that $a_0^{(1)} (k)/\cos^{k-1} \theta = -i \sin \theta$, independent of $k$. Thus, the double sum in Eq.~\ref{eq:l1} is
\begin{align}
& - \sin^2 \theta \sum_{k=1}^{\infty} \sum_{n=k+1}^{\infty} a_0^{(1)} (k) \cos^{n-k-1} \theta\ \widehat{S}_n^- \nonumber \\
= & - \tan^2 \theta \sum_{k=1}^{\infty} \sum_{n=k+1}^{\infty} (-i \sin \theta) \cos^{n} \theta\ \widehat{S}_n^- \nonumber \\
= &  - \tan^2 \theta (-i \sin \theta) \sum_{n=1}^{\infty} \sum_{k=n+1}^{\infty} \cos^{k} \theta\ \widehat{S}_k^- \nonumber \\
= & - \tan^2 \theta (-i \sin \theta) \sum_{k=1}^{\infty} (k-1) \cos^{k} \theta\ \widehat{S}_k^- \nonumber \\
= & - \tan^2 \theta \sum_{k=1}^{\infty} (k-1)\ a_0^{(1)} (k) \cos \theta\ \widehat{S}_k^-
\label{eq:illus}
\end{align}
where we have interchanged the labelling of dummy variables $n$ and $k$ in the third line, and evaluated the sum of geometric series in the fourth. Substituting back into Eq.~\ref{eq:l1} \big(and the more general definition of $\widehat{D}_l^{(1)}$\big), we have
\begin{equation}
a_1^{(1)} (k)  = a_0^{(1)} (k)  \cos \theta\ \big(1 - (k-1) \tan^2 \theta\big).
\end{equation}

We can then apply the same procedure (with Eq.~\ref{eq:wrop1}) to encoding one flying $\ket{\uparrow}_f$ qubit into $\widehat{D}_1^{(1)} = \sum_k a_1^{(1)} (k) \widehat{S}_k^-$ to find, again by linearity, amplitudes for the second 1-spin down-flip distribution 
\begin{align}
\hspace{-4mm} a_2^{(1)} (k) & = a_0^{(1)} (k) \cos^2 \theta\ \times \nonumber \\
& \big(1 - 2(k-1) \tan^2 \theta + \frac{(k-1)(k-2)}{2} \tan^4 \theta \big),
\end{align}
where (as also illustrated in Eq.~\ref{eq:illus}) we have evaluated the following weighted sums of geometric series (by first relabelling the dummy variables to focus on $\widehat{S}_k^-$), 
\begin{align}
\sum_{k=1}^{\infty} \sum_{n=k+1}^{\infty} (k-1) \cos^n \theta\ \widehat{S}_n^- & =  \sum_{n=1}^{\infty} \sum_{k=n+1}^{\infty} (n-1) \cos^k \theta\ \widehat{S}_k^- \nonumber \\
& \hspace{-15mm} = \sum_{k=2}^{\infty} \bigg(\sum_{n=1}^{k-1} (n-1)\bigg) \cos^k \theta\ \widehat{S}_k^- \nonumber \\
& \hspace{-15mm} = \sum_{k=3}^{\infty} \frac{(k-1)(k-2)}{2} \cos^k \theta\ \widehat{S}_k^-. 
\end{align}
Note that the general term of the bracketed series are obtained from evaluation of the weighted geometric series in the previous step (evaluating the lower order distribution).

Now, with the key inductive steps to relabel the dummy variables (to focus on $\widehat{S}_k^-$), and to evaluate the following series (as done above and in Eq.~\ref{eq:illus}),
\begin{equation}
\hspace{-4mm} \sum_{n=l'+1}^{k-1} (-1)^{l'} \frac{1}{l'!}\prod_{m=0}^{l'-1} (l'-m) \frac{1}{l'!}\prod_{j=1}^{l'} (n-j) \equiv \frac{(-1)^{l'+1}}{(l'+1)!}\prod_{j=1}^{l'+1} (k-j),
\end{equation}
we can inductively derive
\begin{align}
\hspace{-3mm} a_l^{(1)} (k) & = a^{(1)}_0(k) \cos^{l} \theta\ \times \nonumber \\
& \hspace{-4mm} \sum_{r=0}^{\text{min}\{l, k-1\}} (-1)^r \bigg(\frac{1}{r!}\prod_{m=0}^{r-1} (l-m)\bigg)\frac{1}{r!}\prod_{j=1}^{r} (k-j) \tan^{2r} \theta \nonumber \\
& = a^{(1)}_0(k) \cos^{l} \theta \sum_{r=0}^{\text{min}\{l, k-1\}} (-1)^r \binom{k-1}{r} \binom{l}{r} \tan^{2r} \theta\  \nonumber \\
&= a^{(1)}_0(k) \cos^{l} \theta\ _2F_1 (1-k, -l; 1; - \tan^2 \theta). 
\label{eq:lgen}
\end{align} 
Note that Eq.~\ref{eq:wrop1} has two terms, the first (and lower order) of which inductively adds to the lower order terms (in $\tan^2 \theta$) in Eq.~\ref{eq:lgen}, to give rise to the bracketed product coefficient concerning $l$. $\square$

\vspace{0.5mm}
In the more compact form of Eq.~\ref{eq:lgen}, $_2F_1(a, b; c; z)$ denotes the {\it Gauss} hypergeometric function~\cite{beals10}, and $a^{(1)}_0(k) \cos^{l} \theta$ renders possible divergence of $\tan^{2r} \theta$ convergent in Eq.~\ref{eq:lgen} (analytic continuation is assumed implicitly here). The expression for the $l^{th}$ collective 1-spin down-flip distribution can also be obtained through a combinatorial argument, as follows. For any fixed down-flip position $k$, and assuming  $k>(l+1)$, the amplitude is a sum of $l+1$ terms , each of which corresponds to a different origin for the $\widehat{S}_{k}^{-}$. In general a spin-up qubit passing along the chain can either cause no spin flips at all, or can move a spin down from a site nearer the front of the chain to one further along it. The $0^{th}$ term results from the situation in which the down spin is initially localized on spin state $k$ and where all $l$ subsequent $\ket{\uparrow}_f$ qubits move along the chain without executing further flips. Each has contributed a factor of $\cos \theta$ due to the exchange interaction and thus an overall factor of $\cos^{l} \theta$ is present in addition to $a^{(1)}_0(k)$. In general, the $r^{th}$ term ($r>0$) occurs when $r$ movements of the initial spin down position occur before that spin down reaches its final position $k$. There are $\binom{l}{r}$ ways of choosing the $r$ qubits which cause the flips from the $l$ total, and $\binom{k-1}{r}$ ways of choosing which $r$ of the $(k-1)$ static spins which precede the $k^{th}$ will hold the spin down at some point before the spin down finally occurs at site $k$. The other terms in the summation come from the fact that each double spin flip (or movement of the spin down location) gives rise to a factor of $(-i \sin \theta)^2$, while losing a factor of $\cos \theta$; in addition, the other $(l-r)$ $\ket{\uparrow}_f$ qubits passed the down-flipped spin without exchanging, and each contributed a factor of $\cos \theta$. Combining these coefficients gives rise to Eq.~\ref{eq:lgen}. 

Decoding this more general memory state can be achieved by injecting successive $\ket{\uparrow}_f$ spins in the decoding direction. After the first such spin passes, the new memory state down-spin amplitude for the $k^{th}$ site, $a'^{(1)}_{l-1} (k)$, results from two possible scenarios: Either this $\ket{\uparrow}_f$ passed the $k^{th}$ site which was already in the down state, without exchange, or it transported the $(k+s)^{th}$ down spin to the $k^{th}$ position. Taking into account of the factors contributed, we have
\begin{eqnarray}
a'^{(1)}_{l-1} (k) &=& a^{(1)}_{l} (k) \cos \theta + (-i \sin \theta)^2 \sum_{s=1}^{\infty} a^{(1)}_{l} (k+s) \cos^{s-1} \theta \nonumber \\
&\equiv& a^{(1)}_{l-1} (k).
\label{eq:rev}
\end{eqnarray}
To establish this last equivalence, we multiply both sides of the following identity (see Appendix II)
\begin{equation}
\hspace{-5.5mm} \frac{_2F_1(a,b;1;z)}{1-z} + \frac{z}{1-z} \sum_{s=1}^{\infty} \frac{_2F_1(a-s,b;1;z)}{(1-z)^s} \equiv \ _2F_1(a,b+1;1;z)
\label{eq:id}
\end{equation}
by $a_0^{(1)} (k) \cos^{l-1} \theta$, and substitute for $a=1-k, b = -l$, and $z = - \tan^2 \theta$; we then obtain $a'^{(1)}_{l-1} (k) \equiv a^{(1)}_{l-1} (k)$ from Eq.~\ref{eq:rev}. Here, $l \in \mathbb{N}$ can be arbitrary. This means that the 1-spin down-flip distributions can be manipulated in both directions, essential for the chain to act as a memory. 

Having established this important feature, we now show that each distribution corresponds to a unique storage mode and the modes are independent, i.e., expressed as the following (discrete) orthonormal condition
\begin{equation}
\sum_{k=1}^{\infty} a^{{(1)}*}_{l'} (k)\ a^{(1)}_{l} (k) = \delta_{l'l}.
\label{eq:orth1}
\end{equation} 
This ensures the unit total probability for each distribution. To establish this discrete orthonormal condition Eq.~\ref{eq:orth1}, we first introduce the normalized {\it Meixner} polynomials~\cite{beals10} (with $j, x$ integers)
\begin{align}
M'_j (x; \mu, \nu)  :&=\nu^{\frac{j}{2}} \ _2F_1\left(-j, -x; \mu; 1 - \frac{1}{\nu}\right) \nonumber \\
& \equiv \nu^{\frac{j}{2}}\  _2F_1\left(-x, -j; \mu; 1 - \frac{1}{\nu}\right) 
\label{eq:mex}
\end{align}
where the equivalence comes from the symmetry of the hypergeometric function in its first two arguments. Note that the different normalization is present since we are only summing over $x \in \mathbb{N}$ (instead of $\mathbb{Z}$). The orthonormality condition for the Meixner polynomials states
\begin{equation}
\sum_{x=0}^{\infty} M'_j (x; \mu, \nu) M'_{j'} (x; \mu, \nu) \omega(x; \mu, \nu) = \delta_{jj'} 
\label{eq:orthmex}
\end{equation}
where the discrete weight $\omega(x; \mu, \nu) = (1-\nu)^{\mu} \frac{(\mu)_x}{x!} \nu^{x}$~\cite{beals10}. Setting $x=k-1, j=l, \mu=1, \nu=\cos^2 \theta$, and substituting Eq.~\ref{eq:mex} into Eq.~\ref{eq:orthmex}, we have
\begin{align}
\hspace{-2mm} \sum_{k=1}^{\infty} \ _2F_1(1-k,-l;1;-\tan^2 \theta)\ & _2F_1(1-k,-l';1;-\tan^2 \theta) \nonumber \\
&\hspace{-12mm} \times \sin^2 \theta \cos^{2(k+l-1)} \theta = \delta_{ll'} 
\end{align}
which is exactly the desired condition Eq.~\ref{eq:orth1}. Note that one special case of the Meixner polynomials are the Krawtchouk polynomials~\cite{beals10}, which have recently been applied to works involving quantum state transfer of a single spin excitation within certain linear, interacting spin chains~\cite{jeugt10}. 

\begin{figure}[h]
\begin{center}$
\begin{array}{cc}
  \subfigure[]{\includegraphics[width=1.6in]{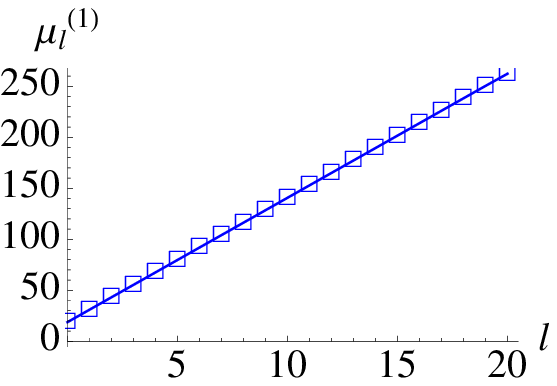}} &
  \subfigure[]{\includegraphics[width=1.6in]{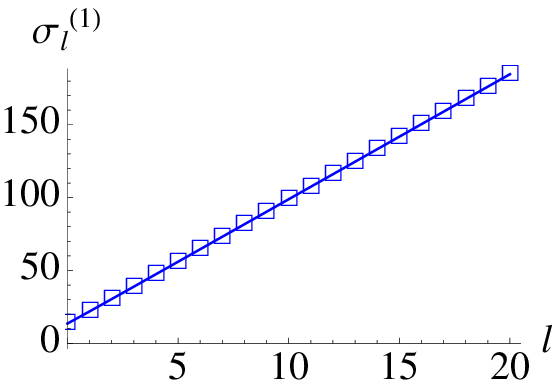}}  
\end{array}$
\caption{Plots with a non-special coupling strength $\theta = 0.4$ for (a) the mean positions $\mu_l^{(1)}$ of the down-flip against the number $l$ of subsequent flying $\ket{\uparrow}_f$ qubits encoded into the memory; (b) the corresponding standard deviations $\sigma_l^{(1)}$ against $l$. The linear relationships are also observed for other $\theta$ values.}
\label{fig:mean}
\end{center}
\end{figure}

The independence of these unique storage modes are best illustrated by considering their mean positions $\mu_l^{(1)}$ of the down-flip and the corresponding standard deviations $\sigma_l^{(1)}$, which can be calculated in a similar way to how we found Eq.~\ref{eq:mean}. We find that for any given coupling strength $\theta$, both $\mu_l^{(1)}$ and $\sigma_l^{(1)}$ increase {\it linearly} with $l$; an example is demonstrated in Fig.~\ref{fig:mean}, with a non-special value $\theta = 0.4$, say.

\begin{figure}[h]
\begin{center}
\includegraphics[width=2.2in]{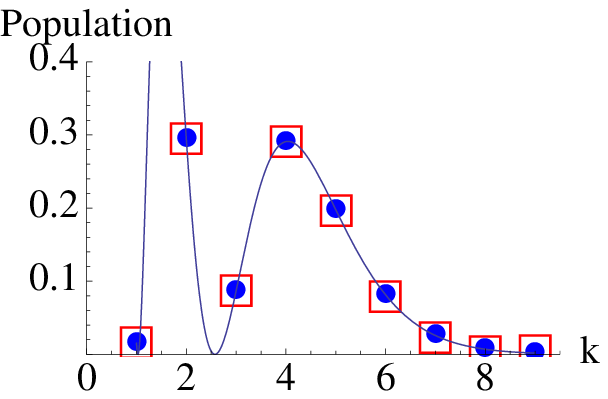}
\caption{Plots of the down-flip distribution $|a_2^{(1)} (k)|^2$ in a chain with $N=9$, and $\theta=1.2$: The discrete plots are from our simulations (``blue circle'' for $|a_2^{(1)} (k)|^2$ and ``red square'' for $|a'^{(1)}_2 (k)|^2$), while the curve corresponds to the analytical solution Eq.~\ref{eq:lgen} (only values for integer $k$ are relevant).}
\label{fig:compare}
\end{center}
\end{figure}

\vspace{1mm}
Finally, we numerically simulate (see Appendix III) the down-flip distributions, by encoding a flying $\ket{\downarrow}_f$ qubit followed by $l$ subsequent $\ket{\uparrow}_f$ into the initially ferromagnetic chain. An example is shown in Fig.~\ref{fig:compare}, and shows agreement with our analytical solutions for $|a_l^{(1)} (k)|^2$.

\section{The (0, 0)$^{th}$ 2-spin down-flip distribution}
To generalize the results, we need to show that the chain can store a spin state that includes an arbitrary number $n$ of down-spins
\begin{equation}
\hspace{-.5mm} \widehat{D}^{(n)}_{(l_1, ..., l_n)} = \sum_{k_1 < ... < k_n} a^{(n)}_{(l_1, ..., l_n)} (k_1, ..., k_n)\ \widehat{S}^-_{(k_1, ..., k_n)}
\end{equation}
where the $k_i$s denote the spin-flip positions and $l_i$ denotes the number of $\ket{\uparrow}_f$ encoded between the $i^{th}$ and $(i+1)^{th}$ flying $\ket{\downarrow}_f$ qubits. This would mean that the chain can store the information from a number of flying qubits, each of which was originally either $\ket{\uparrow}_f$ or $\ket{\downarrow}_f$ and can be further retrieved by the aforementioned decoding mechanism. By linearity, it could also store any superposition, which would confirm its status as a true quantum memory. We, however, are unable to prove the general case due to the increasing complexity of the analytical solution (with large numbers of parameters $l_i$s and $k_j$s). 

Here, we derive the analytical expression for the $(0, 0)^{th}$ collective 2-spin down-flip distribution 
\begin{equation}
\widehat{D}^{(2)}_{(0,0)} = \sum_{k_1 < k_2} a^{(2)}_{(0,0)} (k_1,k_2)\ \widehat{S}^-_{(k_1,k_2)}
\end{equation}
after having encoded only two $\ket{\downarrow}_f$ in the chain. Going through step-by-step we find that the total state of the system, after encoding one flying $\ket{\downarrow}_f$ qubit into a chain with state $\widehat{S}_k^- \ket{F}_c$, becomes
\begin{equation}
\ket{\uparrow}_f \bigotimes (-i \sin \theta) \sum_{k'=2}^{\infty} \cos^{k'-2} \theta\ \widehat{S}_k^- \widehat{S}_{k'}^- \ket{F}_c \hspace{5mm} \text{if}\ k=1,
\end{equation}
or $\forall$ finite $2 \le k \ll N$,
\begin{align}
\hspace{-3.5mm} \ket{\uparrow}_f \bigotimes & (-i \sin \theta) \bigg[ \sum_{k'=k+1}^{\infty} \cos^{k'-2} \theta\ \widehat{S}_k^- \widehat{S}_{k'}^- + \sum_{k'=1}^{k-1} \cos^{k'} \theta\ \widehat{S}_k^- \widehat{S}_{k'}^- \nonumber \\
& -\tan^2 \theta\ \sum_{k'=1}^{k-1} \sum_{k''=k+1}^{\infty} \cos^{k'+k''-k} \theta\ \widehat{S}_{k'}^-  \widehat{S}_{k''}^-\bigg]\ket{F}_c. 
\end{align}
Note that by applying a combinatorial argument as illustrated before, one can obtain the same results.

Thus by linearity, after encoding the $\ket{\downarrow}_f$ into $\widehat{D}_0^{(1)} \ket{F}_c$, we have
\begin{align}
\widehat{D}_{(0,0)}^{(2)} = (-i \sin \theta)^2 & \bigg[\sum_{k=1}^{\infty} \sum_{k'=k+1}^{\infty} \cos^{k + k' -3} \theta\ \widehat{S}_{k}^- \widehat{S}_{k'}^-  + \nonumber \\
& \sum_{k=2}^{\infty} \bigg(\sum_{k'=1}^{k-1} \cos^{k + k' -1} \theta\ \widehat{S}_{k}^- \widehat{S}_{k'}^- \nonumber \\
& \hspace{-15mm} - \tan^2 \theta\ \sum_{k'=1}^{k-1} \sum_{k''=k+1}^{\infty} \cos^{k' + k'' -1} \theta\ \widehat{S}_{k'}^- \widehat{S}_{k''}^-\bigg)\bigg].
\end{align}
By manipulating the double and triple summations as before, we find that
\begin{align}
& \widehat{D}_{(0,0)}^{(2)} = (-i \sin \theta)^2 \sum_{k_1=1}^{\infty} \sum_{k_2=k_1+1}^{\infty} \bigg(\cos^{k_1 + k_2 -3} \theta\ + \nonumber\\ 
& \cos^{k_1 + k_2 -1} \theta - (k_2-k_1-1) \tan^2 \theta \cos^{k_1 + k_2 -1} \theta\bigg) \widehat{S}_{k_1}^- \widehat{S}_{k_2}^-.
\end{align}
Therefore, the $(0, 0)^{th}$ collective 2-spin amplitude is
\begin{equation}
\hspace{-4mm} a_{(0,0)}^{(2)} (k_1, k_2) = (-i \sin \theta)^2 \cos^{k_1+k_2-1} \theta\ \big(2-(k_2 - k_1 - 2) \tan^2 \theta\big), 
\label{eq:twoflip}
\end{equation}
as quoted in the main text. We have the total probability $\sum_{k_1<k_2}|a_{(0,0)}^{(2)} (k_1, k_2)|^2 = 1$. The qubits can also be recovered sequentially, as can be shown by going through step-by-step or using a combinatorial argument.

\vspace{1mm}
We have thus shown that the memory can encode and decode $\ket{\uparrow_{f_2} \uparrow_{f_1}}$, $\ket{\uparrow_{f_2} \downarrow_{f_1}}$, $\ket{\downarrow_{f_2} \uparrow_{f_1}}$, and $\ket{\downarrow_{f_2} \downarrow_{f_1}}$; by linearity, the memory can store two qubits of arbitrary state, {\it entangled} or not (and hence qubits of mixed states). We conjecture that the memory chain can store multiple flying qubits of arbitrary states using the described mechanism (see an example of numerical simulations of larger memory states from Fig.~4 in the main text supporting this conjecture).


\section{Variations in coupling strengths}
In the main text, we have shown an example of numerical simulations involving the memory storage and retrieval of four randomly-generated, pure qubits, with the same coupling strength ($\theta^i = \theta\ \forall\ i$). Here, we consider the effects of the different kinetic energy that each qubit may possess, and the memory's tolerance to the resulting coupling strength variations.

\begin{figure}[h]
\begin{center}$
\begin{array}{cc}
  \hspace{-1mm} \subfigure[\ same $\theta^i = 1$]{\includegraphics[width=1.7in]{Figure5.eps}} &
  \subfigure[\ random $\theta^i \in (0.9, 1.1)$]{\includegraphics[width=1.7in]{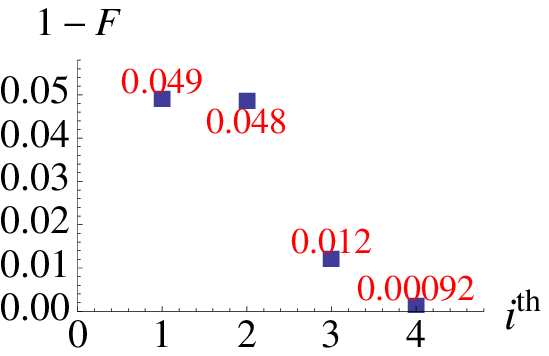}}  
\end{array}$
\caption{Plot for the infidelity $1 - F$ of the retrieved qubit (with phase corrected), relative to the corresponding input $\rho_i$, against the ordinal number $i$ of the inputs. Here, four randomly-generated pure qubits \{$\rho_i$\}$_4$ (same as in the main text) are sequentially encoded into the memory chain $\ket{F}_c$ ($N=9$), and then retrieved one by one in the reverse order without decoherence. $\theta^i$ denotes the coupling strength between the $i^{th}$ qubit $\rho_i$ and each static spin during both encoding and decoding rounds; site-to-site variations are ignored here, and considered later. Note that for (b), ten independent runs of simulations are performed, and for each qubit the average is taken for the phase-corrected fidelity (to discount the imperfect randomness in $\theta^i$ for small number of rounds $n$ and hence possible outliers).}
\label{fig:var1}
\end{center}
\end{figure}

\begin{figure}[h]
\begin{center}$
\begin{array}{cc}
  \hspace{-2mm} \subfigure[\ $\rho_1$]{\includegraphics[width=1.7in]{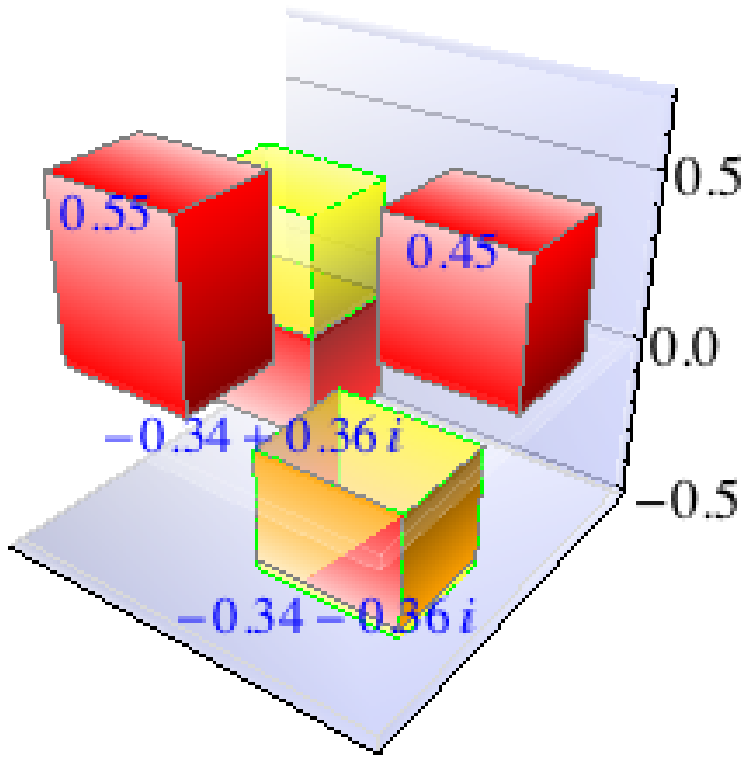}} &
  \subfigure[\ $\rho_1^{ret}$]{\includegraphics[width=1.6in]{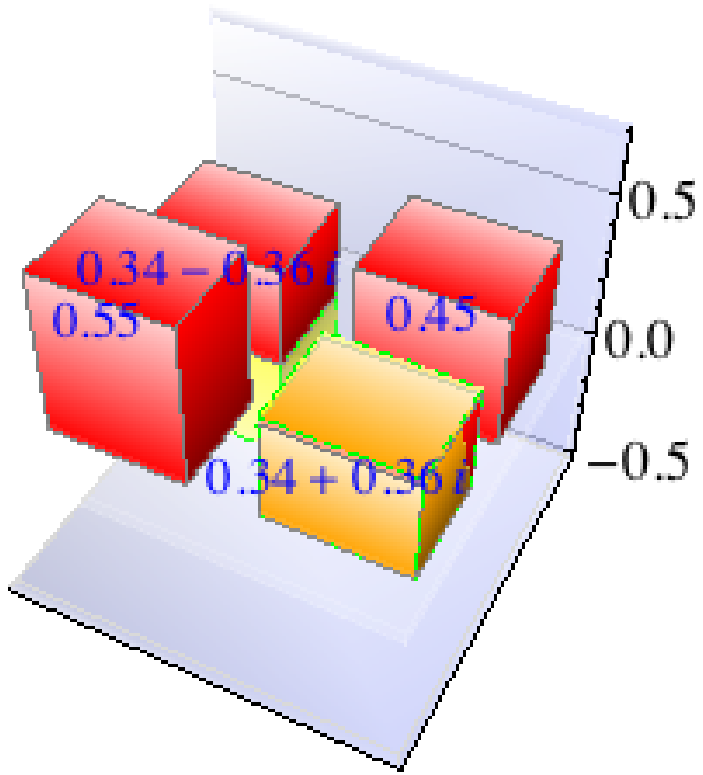}}  
\end{array}$
\caption{Density state tomograms (``red" for real components and ``yellow" for imaginary) constructed for (a) the input $\rho_1$ generated above (i.e., from the main text); (b) the retrieved qubit $\rho_1^{ret}$, by encoding {\it only} $\rho_1$ into the the memory chain $\ket{F}_c$ (with $N=9$) and then read back (without correcting the phase). We have random $\theta_k \in (0.9, 1.1)$ between the qubit and the $k^{th}$ static spin. The infidelity of the retrieved qubit, if phase corrected, is $7.4 \times 10^{-6}$, which is again averaged over the results from ten independent runs of simulations. For comparison, the corresponding infidelity for $\theta_k = 1$ $\forall k$ is $6.1 \times 10^{-6}$.}
\label{fig:var2}
\end{center}
\end{figure}

To separate the effects of finite chain length, we employ the same qubits \{$\rho_i$\}$_4$ generated in the main text and simulate their storage and retrieval from the memory $\ket{F}_c$ with the same length $N=9$, but with random $\theta^i \in (0.9, 1.1)$ for each $i$. Each $\theta_i$ matches for the corresponding decoding and encoding operations of $\rho_i$ (effects of such mismatches were considered in Fig.~\ref{fig:chi}); site-to-site variations are also ignored, but considered below. Fig.~\ref{fig:var1} thus illustrates the effects of small round-to-round variations ($\sim \pm 10$\%) in the coupling strengths, and shows that the memory is robust towards such imperfections as long as the memory has sufficient capacity to store each qubit reliably (i.e., has a large enough $N$).

\vspace{0.5mm}
To consider site-to-site variations in $\theta_k$, we encode and decode only $\rho_1$ into the chain $\ket{F}_c$ with $N=9$, with variable coupling strengths $\theta_k \in (0.9, 1.1)$ between the qubit and $k^{th}$ static spin. Note that again $\theta_k$ matches for the decoding and encoding operations for the same position $k$. In this case, we found the infidelity (averaged over ten independent runs) between $\rho_1^{ret}$ and $\rho_1$ to be $\sim 10^{-6}$, similar to that in the case of $\theta_k = 1$ $\forall k$ (see Fig.~\ref{fig:var2}). Thus, the memory can also tolerate small site-to-site variations ($\sim \pm 10$\%) in the coupling strengths $\theta_k$.

\section{Potential architectures}
In the main text, we have discussed the semiconductor photonic waveguide system~\cite{ciccarello08}, suitable as a candidate for our memory system. We have also suggested the more speculative optical lattice, which provides the advantage of tunable exchange couplings by controlling the laser fields~\cite{duan03}. A further alternative could be a solid-state system of electrons moving in a gated semiconducting quantum wire. The memory is a chain of QDs, each accommodating an electron spin. A flying qubit is manifested in the spin of a moving electron. A drawback of this system is that the flying-static qubit-spin interaction is of the Heisenberg type~\cite{gunlycke06}, and thus one needs to work with $\theta \simeq \frac{\pi}{2}$. The storage of quantum information is between the same species, and is localised in the memory.

It remains a speculation (and requires further investigation to see) whether our proposed memory system could be extended to perform full quantum computation~\cite{tordrup08, wesenberg09, ping12}.

\section{Appendix I}
Calculations of the mean and standard deviation in Eq.~\ref{eq:mean} involves evaluating series of the following forms
\begin{align}
\sum_{k=1}^{\infty} k y^{k-1} & \equiv \frac{d}{dy} \big(\sum_{k=1}^{\infty} y^{k}\big) = \frac{d}{dy} \bigg(\frac{y}{1-y}\bigg)  \nonumber\\
& = \frac{1}{(1-y)^2}, \tag{A1}
\end{align}
\begin{align}
\sum_{k=1}^{\infty} k^2 y^{k-1} & \equiv \sum_{k=1}^{\infty} (k+1)k y^{k-1} - \sum_{k=1}^{\infty} k y^{k-1} \nonumber \\
& \equiv \frac{d^2}{dy^2} \big(\sum_{k=1}^{\infty} y^{k+1}\big) - \frac{1}{(1-y)^2} \nonumber\\
& \equiv \frac{d^2}{dy^2} \bigg(\frac{y^2}{1-y}\bigg) - \frac{1}{(1-y)^2} \nonumber\\
& = \frac{2}{(1-y)^3} - \frac{1}{(1-y)^2} \nonumber \\
& = \frac{1+y}{(1-y)^3}. \tag{A2}
\end{align}
We then set $y = \cos^2 \theta$ for the two final formulae, and substitute back into Eq.~\ref{eq:mean} for evalutions.

\section{Appendix II}
In order to show that $a'^{(1)}_{l-1}$ (Eq.~\ref{eq:rev}) is the same as $a^{(1)}_{l-1}$, we applied the identity stated in Eq.~\ref{eq:id}:
\begin{equation*}
\hspace{-5.5mm} \frac{_2F_1(a,b;1;z)}{1-z} + \frac{z}{1-z} \sum_{s=1}^{\infty} \frac{_2F_1(a-s,b;1;z)}{(1-z)^s} \equiv \ _2F_1(a,b+1;1;z).
\end{equation*}
To prove this, we work with the {\it Euler}'s integral representation for the hypergeometric function
\begin{equation}
_2F_1(a,b;1;z) = \frac{1}{B(b,1-b)} \int_0^1 \frac{t^{b-1} (1-t)^{-b}}{(1-zt)^{a}} dt \tag{A3}
\label{eq:int}
\end{equation}
for $|z| \le 1$, and apply analytic continuation to other values of $z$. Here, $B(p, q) = \Gamma(p) \Gamma(q) / \Gamma(p+q)$ is the {\it beta} function, where $\Gamma(p+1) = p \Gamma(p)$ is the {\it gamma} function~\cite{beals10}. We see that
\begin{align}
B(b, 1-b) & = \frac{\Gamma(b)\Gamma(1-b)}{\Gamma(1)} = \frac{\frac{1}{b}\Gamma(b+1) (-b) \Gamma(-b)}{\Gamma(1)} \nonumber \\
& = - B(b+1, -b)  \tag{A4}.
\label{eq:beta}
\end{align}
Thus, by substituting both Eqs.~\ref{eq:int} and~\ref{eq:beta} back into Eq.~\ref{eq:id} (and multiplying it both sides by $(1-z) B(b, 1-b)$), our task reduces to proving that
\begin{align}
& \int_0^1 \frac{t^{b-1} (1-t)^{-b}}{(1-zt)^{a}} dt + \sum_{s=1}^{\infty} \frac{z}{(1-z)^s} \int_0^1 \frac{t^{b-1} (1-t)^{-b}}{(1-zt)^{a-s}} dt \nonumber \\
& \equiv - (1-z) \int_0^1 \frac{t^{b} (1-t)^{-b-1}}{(1-zt)^{a}} dt, \tag{A5}
\label{eq:int2}
\end{align}
where the extra minus sign on the right hand side comes from Eq.~\ref{eq:beta}. Evaluating the summation in Eq.~\ref{eq:int2}, we have
\begin{align}
&\sum_{s=1}^{\infty} \frac{z}{(1-z)^s} \int_0^1 \frac{t^{b-1} (1-t)^{-b}}{(1-zt)^{a-s}} dt \nonumber \\
& = \int_0^1 z \bigg[\sum_{s=1}^{\infty} \bigg(\frac{1-zt}{1-z}\bigg)^s\bigg] \frac{t^{b-1} (1-t)^{-b}}{(1-zt)^{a}} dt \nonumber \\
& =  \int_0^1 \cancel{z} \bigg(\frac{1-zt}{\cancel{z}(t-1)}\bigg) \frac{t^{b-1} (1-t)^{-b}}{(1-zt)^{a}} dt.  \tag{A6}
\end{align}
Hence, the left hand side of Eq.~\ref{eq:int2} is equivalent to
\begin{align}
& \int_0^1 \bigg(1 - \frac{1-zt}{1-t}\bigg) \frac{t^{b-1} (1-t)^{-b}}{(1-zt)^{a}} dt \nonumber \\
& = -(1-z) \int_0^1\frac{t}{1-t} \frac{t^{b-1} (1-t)^{-b}}{(1-zt)^{a}} dt, \tag{A7}
\end{align}
which is exactly the right hand side of Eq.~\ref{eq:int2}. $\Box$

\vspace{5mm}
\section{Appendix III}
The time evolution operator for the interaction between a flying qubit and a static spin can be expressed as in Eq.~2 in the main text. However, each flying qubit interacts with the static spins sequentially; in order to numerically simulate the evolution for the state of the whole system $\ket{\psi_f, \Phi_c}$, we apply the write and read operators,
\begin{align}
U_{write} &= U_{fN} ... U_{f2} U_{f1} \nonumber \\
U_{read} &= U_{f1} U_{f2} ... U_{fN} \tag{A8}
\end{align} 
where each $2^{N+1} \times 2^{N+1}$ time evolution matrix $U_{fi}$ describes the interaction between the flying qubit and the $i^{th}$ static spin under the standard basis $\{f, s_1, s_2, ..., s_N\}$):
\begin{equation}
\hspace{-5mm} U_{fk}=\left( {\begin{array}{cccc|cccc}
\mathbb{1}_{2^{N-k}} &  &  & \mathbb{0}_{2^{N-k}} &  &  &  &  \\
 & \mathbb{C}_{2^{N-k}} &  &  & \mathbb{S}_{2^{N-k}} &  &  &  \\
 &  &  \ddots &  &  & \ddots &  &  \\
\mathbb{0}_{2^{N-k}} &  &  & \mathbb{C}_{2^{N-k}} &  &  & \mathbb{S}_{2^{N-k}} & \\
\hline
 & \mathbb{S}_{2^{N-k}} &   &   & \mathbb{C}_{2^{N-k}} &  &  & \mathbb{0}_{2^{N-k}} \\
 &  & \ddots &   &   & \ddots &  & \\
 &  &  & \mathbb{S}_{2^{N-k}} &  &  & \mathbb{C}_{2^{N-k}} & \\
 &  &  &  & \mathbb{0}_{2^{N-k}} &  &  & \mathbb{1}_{2^{N-k}} \\
\end{array}}\right). \tag{A9}
\label{eq:block}
\end{equation}
where $\mathbb{1}_{m}$ is the $m \times m$ identity matrix, $\mathbb{C}_{m} = \cos \theta_k$ $\mathbb{1}_{m}$ while $\mathbb{S}_{m} = -i \sin \theta_k$ $\mathbb{1}_{m}$; both $\mathbb{0}_m$ and blank fields correspond to blocks of 0s. The dots represent the {\it alternating} repetitive patterns (along the diagonal in the upper left and lower right blocks these are $\mathbb{1}_m$, $\mathbb{C}_m$, $\mathbb{1}_m$, $\mathbb{C}_m$ ...; along the diagonals in the lower left and upper right blocks these are $\mathbb{S}_m$, $\mathbb{0}_m$, $\mathbb{S}_m$, $\mathbb{0}_m$ ...), and there are an equal number of blocks of $\mathbb{1}_m$'s, $\mathbb{C}_m$'s and $\mathbb{S}_m$'s. This is readily obtained by considering the exchange processes between the flying qubit and the $k^{th}$ static spin, regardless of the state of other spins (the alternating repetitive pattern due to spins $s_1$ to $s_{k-1}$ and the size of each block due to $s_{k+1}$ to $s_N$). Note that each $U_{fk}$ is symmetric, and centrosymmetric (symmetric about the centre point and a property that is closed under matrix multiplications)~\cite{liu03}. 

When the whole system is restricted to at most one excitation (as we did in the main text for considering decoherence), the state space consisting of more than one excitations is never accessed and can thus be ignored for the purpose of simulation. In this way, the dimension $(N+2)\times(N+2)$ of the density matrix (with only zero or one excitation) grows linearly with the number $N$ of chain spins; simulations of very long chains can then be done conveniently.

\bibliographystyle{apsrev}